# Criticality in a simple model for Earth's magnetic field reversals

A. R. R. Papa[1,2], M. A. do Espírito Santo[1,3], C. S. Barbosa[1], D. Oliva[1],

[1]*Observatório Nacional, General José Cristino 77, 20921-400, Rio de Janeiro, RJ, Brazil.*
[2]*Universidade do Estado do Rio de Janeiro, São Francisco Xavier 524, 20559-900 RJ, Rio de Janeiro, Brazil.*
[3]*Instituto Federal do Rio de Janeiro, Antônio Barreiros 212, 27213-100, Volta Redonda, RJ, Brazil.*

We introduce a simple model for Earth's magnetic field reversals. The model consists in random nodes simulating vortices in the liquid part of the core which through a simple updating algorithm converge to a self organized critical state, with inter-reversal time probability distributions functions in the form of power laws (as supposed to be in actual reversals). It should not be expected a detailed description of reversals. However, we hope to reach a profounder knowledge of reversals through some of the basic characteristic that are well reproduced.

One of the most rooted believes among common people is that the compass needle point to the North, the Earth has a magnetic field that imposes this behavior. However, it not has been like that during the whole Earth's history. In different geological epochs the needle would have pointed in the opposite direction. The magnetic field of the Earth has changed its direction many times during the last 160 million years (Myr). Each of these changes is called a reversal. Their existence has been documented through a large number of experiments in the oceanic floor and other locations where the magnetization at the time of their formation has been recorded in rocks. Reversals are, together with earthquakes, some of the more astonishing events on Earth.

The interest in the study of geomagnetic reversals comes from the contribution they can add to the comprehension of important processes in Earth's evolution, which includes from the internal dynamics of the planet to processes involving the evolution of living organisms. The Earth is continuously bombarded by charged particles mainly born at the Sun. Large fluxes of those particles are incompatible with life. Fortunately, only a tiny fraction of them hits the Earth's surface. The most part is deviated by the geomagnetic field. However, during reversals the field intensity decreases drastically which means a more intense bombarding. The so-called magnetic shield is reduced. Those increases in the flux of particles could have interfered in life cycles of the planet, including mass extinctions [1].

Periods between geomagnetic reversals (called chrons) seem to present power law distribution functions, which can be the signature of some self-organized critical system as the mechanism of their source [2]. Among other mechanisms capable of producing power laws we can mention the superposition of some distributions [3], critical systems [4] and non-extensive versions of statistical mechanics [5], on which we will not further explore upon here.

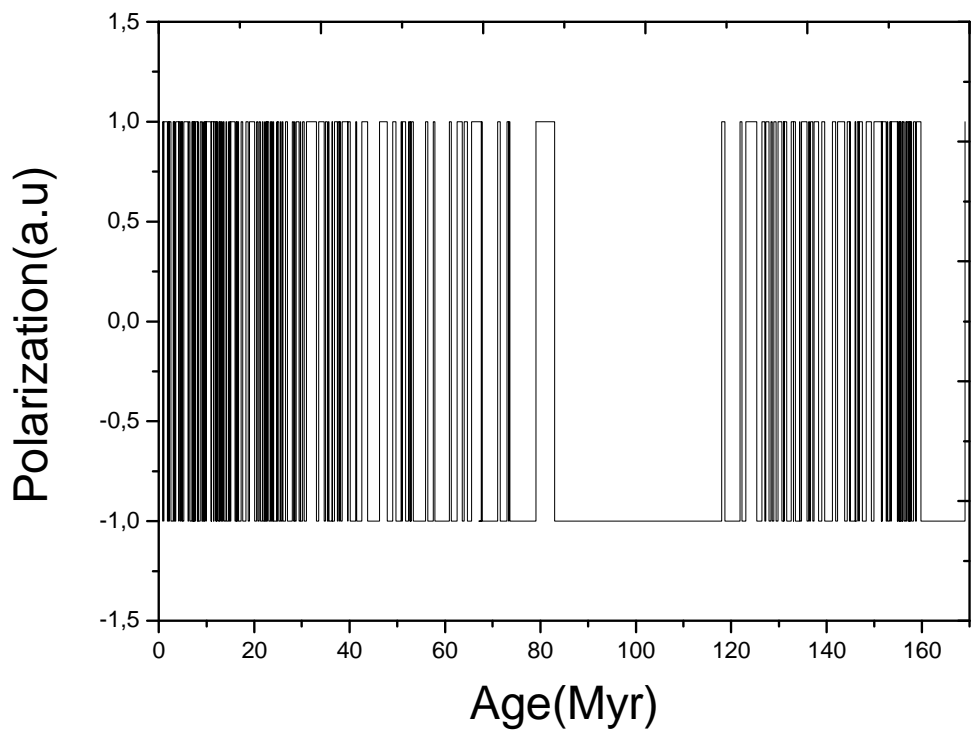

Figure 1.- A binary representation of geomagnetic reversals from around 160 Myr ago to our days. We arbitrarily have assumed -1 as the current polarization. There was a particularly long period with no reversals from 120 to 80 Myr. The period from 165 Myr to 120 Myr was very similar to the period from 40 to 0 Myr (in both, the number of reversals and the average duration of reversals). The data was obtained from Cande and Kent [6,7].

In Fig. 1 we present the sequence of geomagnetic reversals from 160 Myr to our days. The data was obtained from the most recent and complete record that we have found [6,7]. Fig. 2 presents the distribution of periods between reversals which follows a power law,

$$f(t) = c \cdot t^d \tag{1}$$

where f (t) is the frequency distribution of periods between consecutive reversals, c is a proportionality constant and d is the exponent of the power-law (and also the slope of the graph in log-log plots). For the present case we have approximately -1.5 as slope value.

The magnetic field needs a finite time to change its direction (the time between the moment at which the dipolar component is no more the main one and the moment at which it is again the main one but building up in the opposite direction). This time is in the average around 5000 years. It is small if compared to the smallest interval between reversals already detected (10,000 years) and to the average interval between reversals. There are clustered periods of high activity from 40 Myr ago to our days and during the period 165-120 Myr, a period of low activity between 80 and 40 Myr ago and a period of almost null activity from 120 to 80 Myr.

The sequence of Earth's magnetic field reversals seems to be a non-equilibrium process as can be inferred from Fig. 1 where the average time interval between successive reversals seems to increase with geological time (at least from 80 Myr to our days). Unfortunately, the reversal series is unique and relatively small, there is no other similar.

Contrary to the record of reversals (known from 160 Myr up to nowadays) the record of magnetic field intensities exists just for around 10 Myr [8]. Fig. 3 shows the distribution of Earth's dipole values for the last 10 Myr. The distribution of dipole values for actual reversals (Fig. 3) follows a function between a bi-normal and a bi-log-normal distribution. However, the number of experimental points is also small and there will certainly be changes in these facts when new measurements become available.

Geomagnetic reversals have called a great attention since their discovery and this interest remains nowadays. Works devoted to the study of the time distribution of geomagnetic reversals include, among many others, statistical studies on reversals distributions and correlations [2,9-11], modeling of the problem [12-14] and experimental studies [6,7,15-17]. Gaffin [9] pointed out that long-term trends and non-stationary characteristics of the record could hamper a formal detection of chaos in geomagnetic reversal record.

It is our opinion that actually it is very difficult to detect in a consistent manner that geomagnetic reversals present any systematic characteristic at all, without mattering which this characteristic could be (including chaos). Its nature remains an open question.

Here we profit from all the accumulated experience mentioned before. However, our model follows lines quite different from theirs. Without appealing to a detailed description of edy vortices and their mutual interaction in the external core, we represent them by random numbers. The weakest vortices are systematically removed and substituted by new ones. So are the numbers in the neighborhood of the weakest.

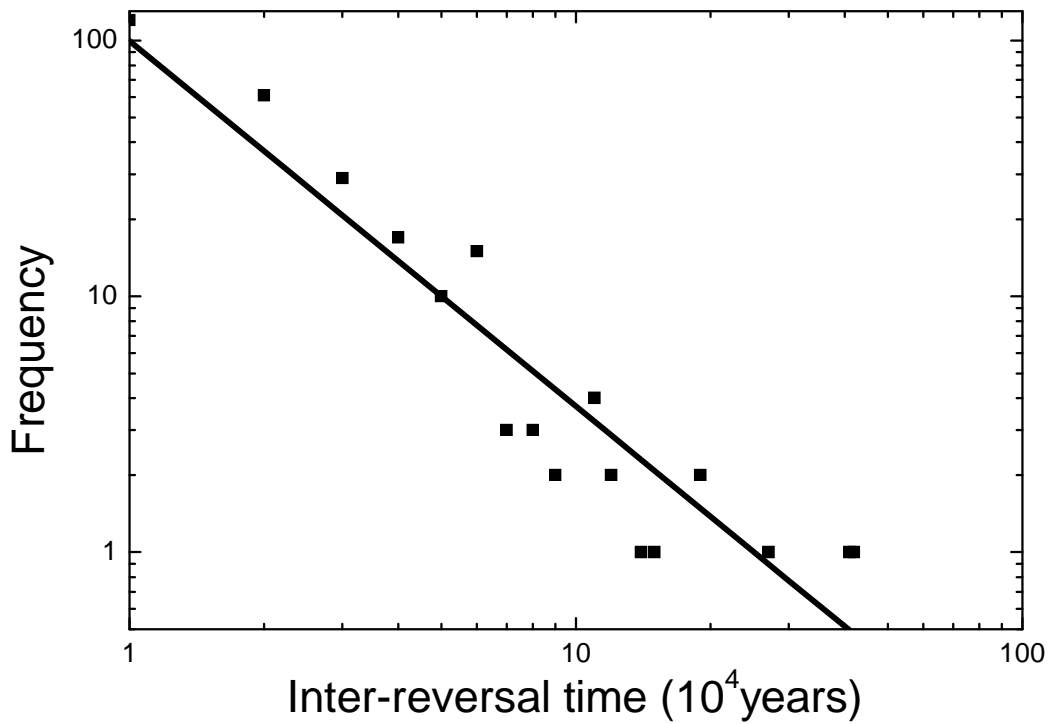

Figure 2.- Frequency distribution of periods between reversals from 160 Myr to our days. The slope of the solid line is around -1.5. The data was obtained from Cande and Kent [6,7].

The rest of the paper is organized as follows: first, we present the model, later on we present the results of our simulations as well as a comparative study with experimental results and a discussion on their possible connection with previous works on the statistics of geomagnetic reversals. Finally, we present our conclusions and some possible trends for future works.

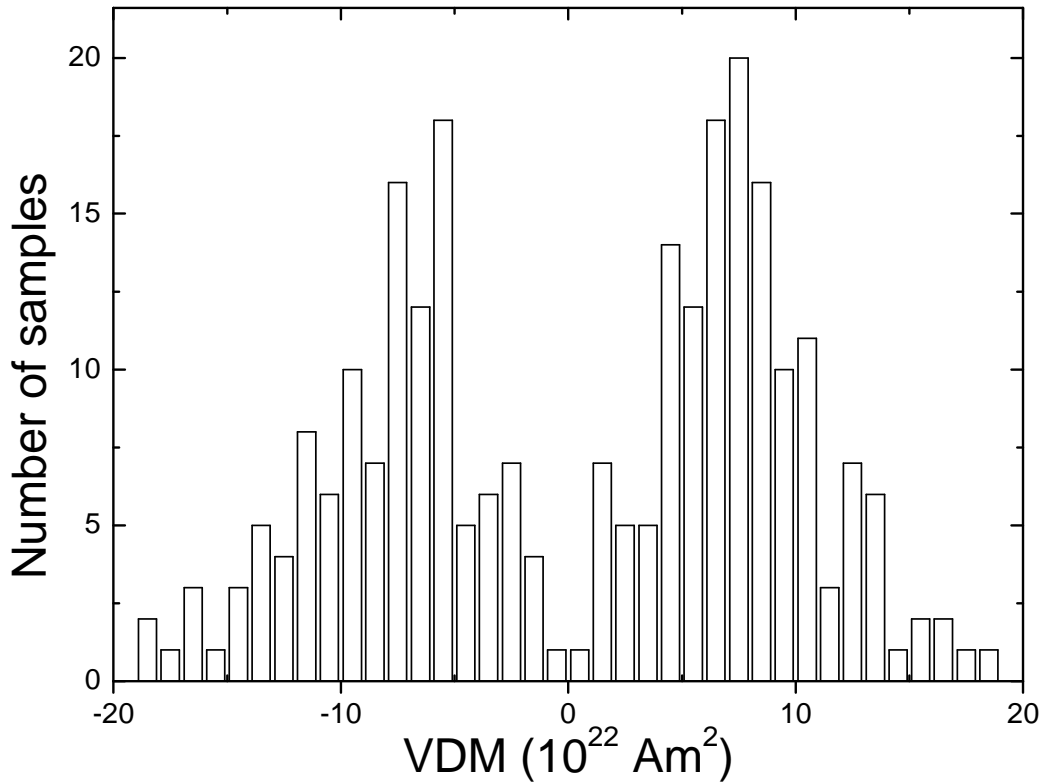

Figure 3.- Distribution of virtual dipole moment (VDM) values for samples not older than 10 Myr. Negative values of VDM correspond to reverse polarization and are represented by 119 samples. Positive values of VDM represent normal polarization and are represented by 142 samples. The data was obtained from Kono and Tanaka [11].

We simulate the Earth's liquid core and the electric current structures on its volume by nodes distributed on an LxL square lattice, where L=100, 200 and 300. This gives different numbers of nodes on each simulation from where we have checked for finite size effects. We explore in sets of equally spaced points at the Earth's equator. To each of these nodes we have initially assigned a random value between -1 and 1 to simulate both, the accumulated magnetic energy at each of the simulated positions and the magnetic moment orientation. We have looked then for the lowest absolute value through the whole system and changed it and its four nearest neighbors by new random values between -1 and 1 (this defines our time step). With this we simulate a more or less continuous energy flux to the core bulk (this is the reason to pick the lowest value) and the possible absorption of smaller vortices. In this way we also simulate the creation of new vortices. At the same time, the assignment of new random values, to the lowest in absolute and its neighbors, works as a continuous release of energy out of the system.

This process is repeated several times (usually between $10^6$ and $10^8$ times) to obtain stationary distributions for the quantities we are interested in.

We define the magnetization M for the system as:

$$M = \frac{1}{N}\sum_{i=1}^{N} s_i \qquad (2)$$

where the sum runs over all the nodes and N = LxL is the total number of nodes. It can take values between -1 and 1 (corresponding to all nodes in the -1 value and to all nodes in the 1 value, respectively).

Constructed in the way we have done, our model qualifies as a Bak-Sneppen one [1]. The Back-Sneppen model probably is the simplest model presenting self-organized criticality, i.e., the tendency to a stationary critical state without necessity of a fine tuning.

The Bak-Sneppen model is a general model that has found applications in a large number of fields among which we can mention evolution [1], the brain [18], the cosmic rays spectrum [19], X-rays bursts at the Sun's surface [20] and flares [21].

Many scientific efforts have been devoted to characterize the Bak-Sneppen model from several points of view. Examples of them are: its correlations from detrended fluctuation analysis [22], damage spreading on it [23], and its behavior under reduction to near zero dimension [24].

Note that within the simplified model here introduced it should not be expected a detailed description of the system, but just some specific details and, in particular, the class of universal behavior displayed by the real system that the model represents.

Beginning with an arbitrary distribution of accumulated magnetic energy at each node, the subsequent activity will be completely uncorrelated in space and time. But as times goes by (and then, the mean accumulated energy increases in absolute value as a consequence of selecting and changing the lowest absolute values) it will be more and more likely that near neighbors are consecutively changed. After a transient the system reaches a steady state characterized by a well-like distribution for the accumulated energies and an inverse V-shape distribution for the lowest nodes (see Figures 4 and 5). The distribution of lower energies vanishes above + $E_c$ and below - $E_c$, while the distribution of node values is practically nule between these values and presents a plateau above $E_c$ and below –$E_c$.

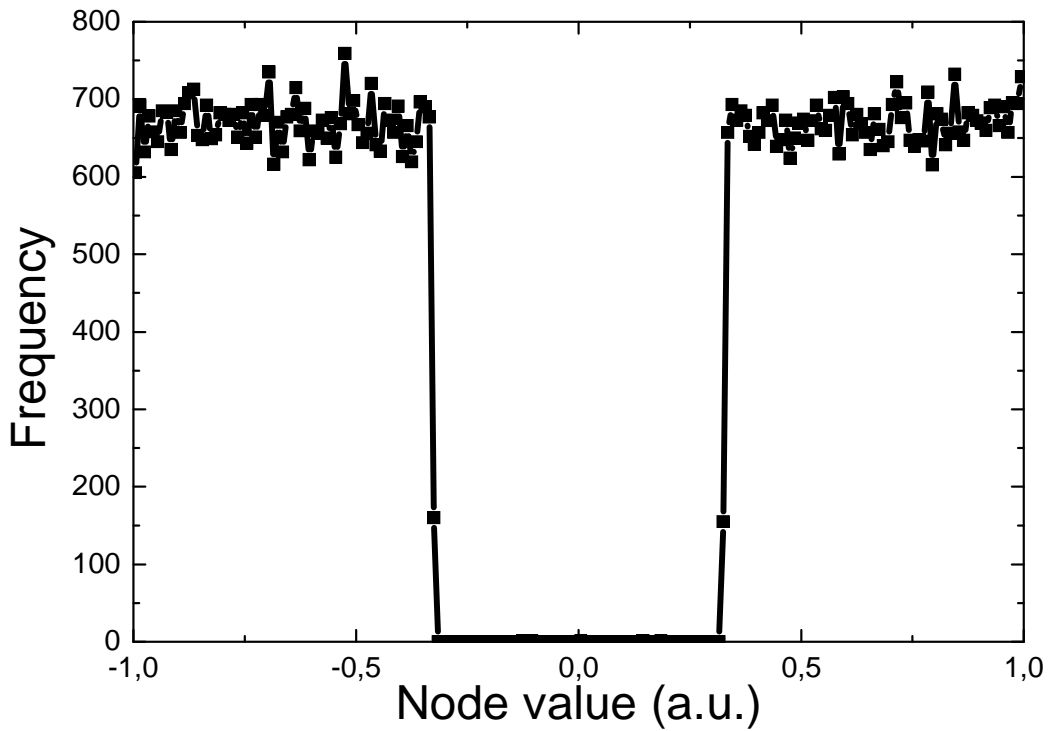

Figure 4.- Distribution of the nodes values at the stationary state. It has a well-like form with vertical walls at $E_c = \pm 0.35$ approximately.

As mentioned before, in the stationary state the events become correlated also in space. In actual reversals this correlation could be associated to jerks (linear temporary trends that appear in the global geomagnetic field but that seldom are measured at different times or filled in some locations but no in others) but we have not done a systematic study on this correlation for the present work. However, it will be the subject of forthcoming works because the power law it follows (not shown) is one of the fingerprints of the critical state on which we believe the system we are simulating is. As in the rest of the simulations the result does not depend on the initial conditions indicating that the critical state is a global attractor for the dynamics, hence it is self-organized.

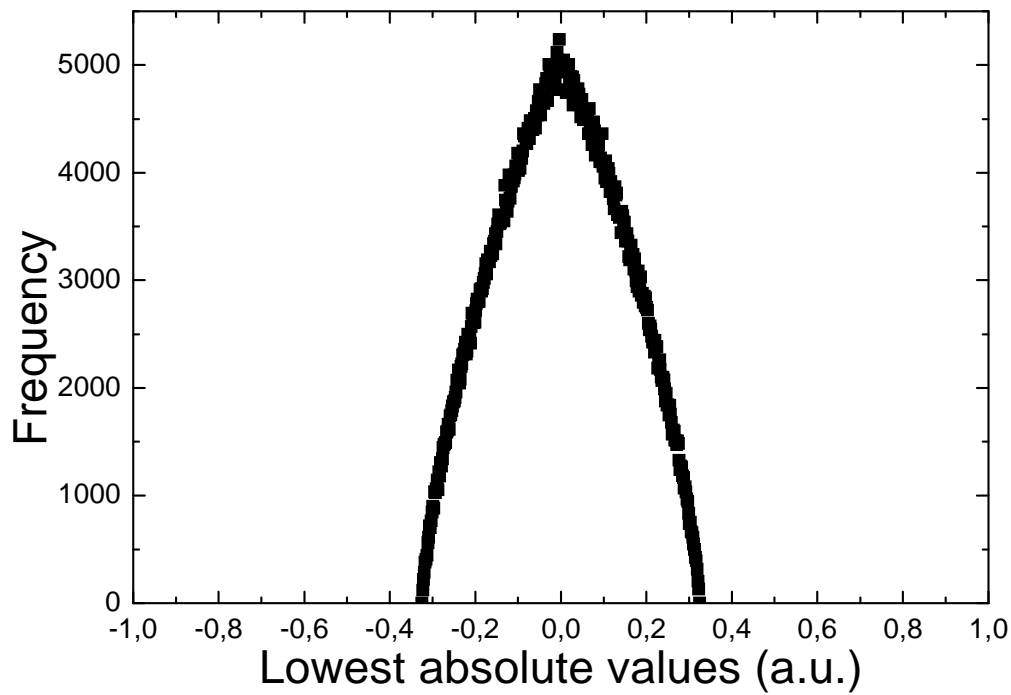

Figure 5.- Distribution of the lower absolute values for nodes at the stationary state. It has an inverse V-like shape with vertical walls at $E_c = \pm 0.35$ approximately.

In Figure 6 we present the dependence of the magnetization on time for a short run. Note that in this particular case there is a preponderancy of positive values for the magnetization.

However, positive and negative values are equivalent. The important fact is that the model presents reversals.

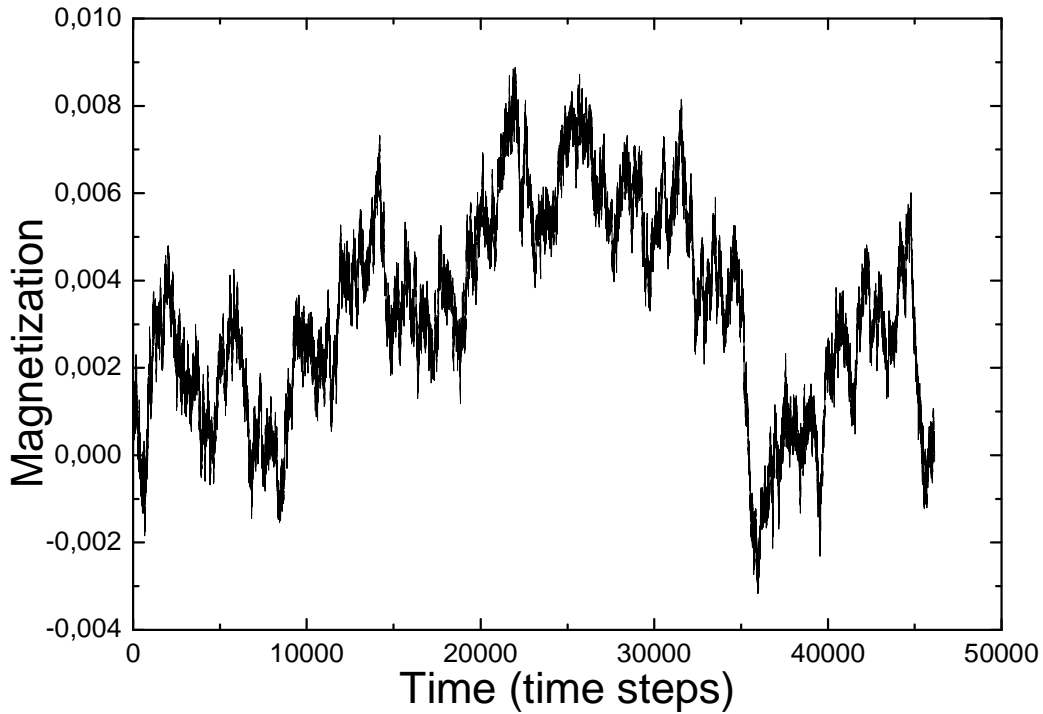

Figure 6.- Magnetization (see text) versus time for a 100x100 system during a short simulation (~45000 time steps).

Figure 7 presents the distribution of inter-reversals times, i.e., the distribution of times during which the magnetization keeps its sign unaltered (chrons in the geomagnetic vocabulary). It follows a power-law with a slope ~ -1.6, very close to the value ~ -1.55 accepted for actual reversals.

Finally, Figure 8 presents the distribution function of nodes values for the data presented in Figure 6. The asymmetry is clear between positive and negative values of magnetizations. Actually, it seems to be not a stationary distribution function because for finite systems there is always a probability for the system to enter a long period of time with the same magnetization sign, able to change any supposedly stationary distribution.

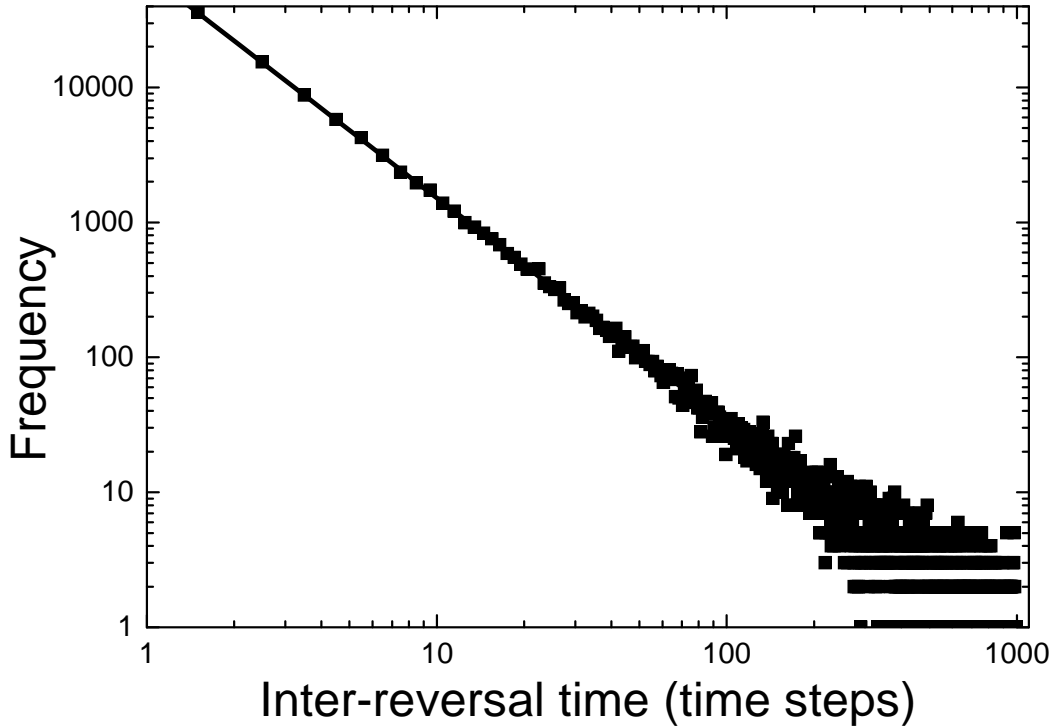

Figure 7.- Distribution of time between consecutive reversals (both from positive to negative and from negative to positive). The slope is ~ -1.6.

It is worth to mention that in our model we are interested in sign changes of magnetization. This contrasts with previous uses of Bak-Sneppen-like models where the important characteristics were avalanches and their probability distribution functions. Avalanches in those cases are often defined as periods of continuous activity (changes in node values) above or below a given threshold. The avalanche picture could, however, play a role also in our model: the definition of reversals just as inversions in the geomagnetic field (in our case, change of sign in magnetization) is not a very rigorous one, because the complexity of the geomagnetic field and their structure around the Earth globe. A more spread definition associates reversals to the changes in sign of the $g_1^0$ coefficient of the representation of the geomagnetic field in spherical harmonics. There are other components (quadrupolar, for example) that can contribute relevantly to the field. However, even this more complex definition doesn't fulfill specific requirements for the Earth. Depending on sites where samples are collected to analyze reversals they present magnetizations in the opposite direction of samples coming from other sites. To permit that the magnetization present the same direction around the whole Earth is normally required a given time. This is the reason to only consider reversals those with a certain stability which translate as those which last for a long enough time. The rest are called excursions. That vision would permit the application of avalanche-like studies in our model by either consider reversals only those periods of time with continuous activity above a given threshold (in Fig. 6, for example, let us say, above 0.2 and below -0.2) or by considering reversals only those that last at least for a given predefined time. Those studies are on course.

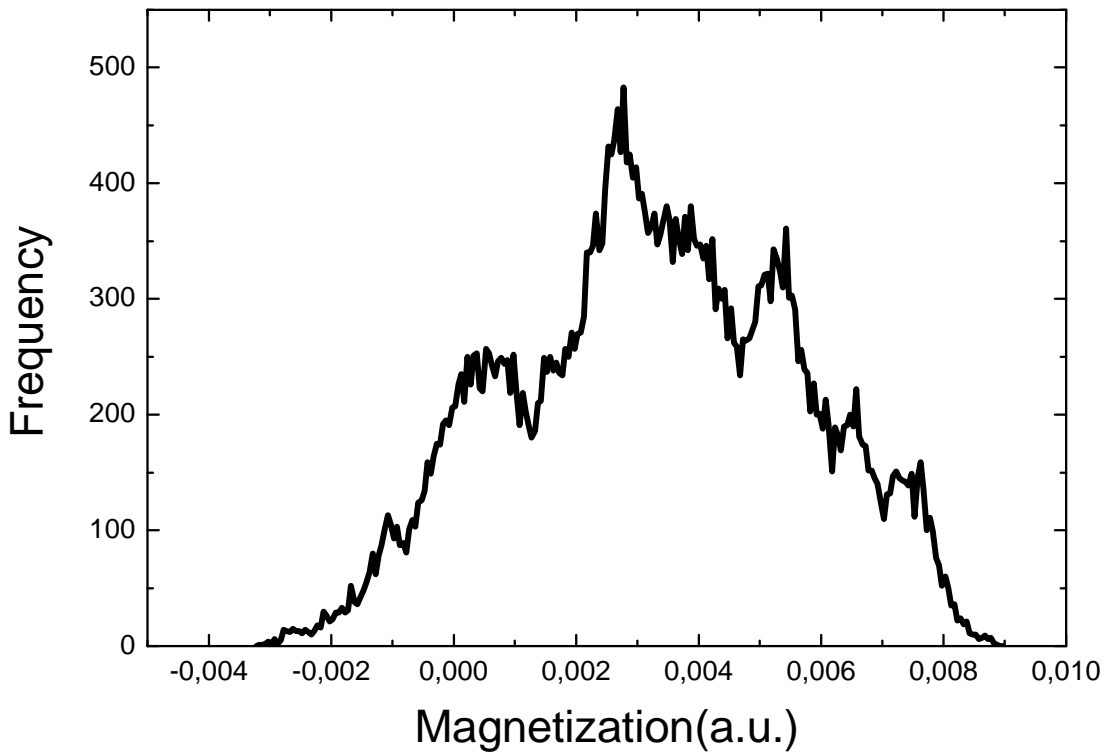

Figure 8.- Distribution function of magnetization values corresponding to Figure 6. Note a preponderancy of positive values in this example (see text). Note also that the magnetization M remains in values much lower in absolute value than the limit value 1.

As concluding remarks let us mention that we have introduced a two dimensional self-organized critical model to simulated Earth's magnetic field reversals, one of the more fascinating geophysical phenomena. The model presents reversals, i.e. changes in the sign of magnetization. We have obtained power law distributions for several relevant quantities, similar to some results of experimental works on reversals. This could imply that the Earth's liquid core could be in a critical state where the greater possible duration of periods between consecutive reversals is limited just by the size of the system (i.e. the volume of the Earth's liquid core). At the best of our knowledge, it is the first SOC model representing reversals and it has also the peculiarity of reaching a single stationary state with two equally probably opposite magnetization states. Elements to be introduced in future works include among others a careful analysis on the spatial distribution of consecutive activity and its possible relations with jerks and a quantitative estimative on the actual ratio between maximum observed magnetization and the maximum magnetization that the systems might support, which corresponds in the real system the model mimics to the whole liquid core of the Earth rotating once by day.

C.S.B., D.O. and A.R.R.P. thank CNPq (Brazilian Science Funding Agency) for MSc, postdoctoral and research fellowships, respectively.